\documentclass[prd,12 pt, preprint,tightenlines,floatfix,showpacs,preprintnumbers,nofootinbib,eqsecnum]{revtex4}

 \usepackage[dvips,final]{graphicx}
  \usepackage{amssymb}
   \usepackage{amsmath}
    \usepackage{amsfonts}
     \usepackage{epsfig}
      \usepackage{bm}% bold math
\usepackage[english]{babel}

\numberwithin{equation}{section}

\begin{document}

\title{The gauge model of quark--meson interactions and the Higgs status of scalar mesons}

\author{V.~Beylin}
\email{vbey@rambler.ru} \affiliation{Research Institute of Physics, Southern
Federal University, Rostov-on-Don, Russia}

\author{V.~Kuksa}
 \email{kuksa@list.ru} \affiliation{Research Institute of Physics, Southern
Federal University, Rostov-on-Don, Russia}

\author{G.~Vereshkov}
\email{gveresh@gmail.com} \affiliation{Research Institute of
Physics, Southern Federal University, Rostov-on-Don, Russia}
\date{\today}

\begin{abstract}
Electromagnetic and strong hadron processes at low energies are considered in the renormalizable model with the spontaneously broken
$U_0(1)\times U(1)\times SU(2)$ gauge symmetry. Calculated radiative
widths of vector mesons and effective couplings $g_{VVS}$ agree with the
experimental data. Residual Higgs degrees of freedom are associated with scalar states $a_0(980)$ and $f_0(980)$ with the degeneration in masses. Two-gamma decays of $\pi^0-$ and $\sigma-$ mesons are analyzed in detail. To provide an "infrared confinement" a cutoff procedure has been also used in calculations.
\end{abstract}

\pacs{12.40Vv, 13.20Jf, 13.25Jx}
\pagenumbering{arabic}\setcounter{page}{1}

\maketitle

\section{The gauge model description}
Low-energy phenomena in the hadron physics are strongly determined by the nonperturbative (NP) interactions.
At the fundamental (QCD) level hadron characteristics are analyzed in the QCD Sum Rules with a NP vacuum condensates. Hadron structure and interactions can be also considered in the effective Lagrangian approach. Starting from the QCD, chiral low-energy expansion of Green functions leads to effective Lagrangian with phenomenological coupling constants \cite{1,2}. Effective chiral hadron model having the Nambu-Jona-Lasinio structure with the quark-meson interactions and incorporating vector mesons as a gauge fields, has been also applied to the hadron reactions studying \cite{3,4}. In fact, quark-meson models contain some input (NP) parameters: effective couplings, which are defined by the (diverged) quark loops. Their regularization fixes some NP scale of the theory, $\Lambda \lesssim 1\, \mbox{GeV}$. Obviously, these models reproduce both nonperturbative and structure effects of the low-energy interactions \cite{5,6}. 

As it will be demonstrated, dynamical symmetry of hadron interactions can be adequately considered in the gauge scheme for hadron and (constituent) quark fields from the very beginning. So, gauge principles, spontaneous symmetry breaking and the Higgs mechanism work well not only at the fundamental quark level, but at the hadron one. In this way, baryon-meson interactions were analyzed in \cite{7}.

The suggested renormalizable quantum field model is the $\sigma-$ model inspired construction, where the fundamental hadron components are reproduced by the constituent quarks and mesons. 
%The linear $\sigma$- model symmetry is used to study meson-meson and (constituent) quark-meson interactions in explicitly chiral form as the %representation of $SU_L(2)\times SU_R(2)$ global group. 
To analyze strong and electromagnetic processes it is sufficient to consider $U_{0}(1)\times U(1)\times SU(2)$ gauge group with the spontaneous symmetry breaking. Then, we localize the diagonal sum of chiral $SU_{L,R}(2)$ subgroups of the global $SU_L(2)\times SU_R(2)$ group only. Due to extra $U(1)$ groups, vector meson dominance is realized at the gauge level as the mixing of initial gauge fields, $\rho$, $\omega$ and $\gamma$ (their chiral partners occur after localization of the total chiral group). Despite of radiative decays of light mesons, the model describes some details of the scalar meson physics. Namely, after the spontaneous breaking of the local symmetry, residual scalar massive degrees of freedom (Higgs fields) reproduce physical scalar states: isotriplet $a_0(980)$ and isosinglet $f_0(980)$. Some properties of the scalar meson spectra are explained in this way. Note, $\sigma-$ meson is associated with $f_0(600)-$ meson \cite{8,9}.  Mesons $\sigma$ and $\pi$ manifest themselves as real external fields, and work also as an effective virtual components of internal hadron vacuum.  Their occurence in loops can be understood as an excitation of the quark sea in a hadron.

The initial model Lagrangian can be written as follows:
\begin{eqnarray}\label{E:8}
&L=i \bar{q} \hat{D} q - \varkappa \bar{q} (\sigma +i\pi^a\tau_a \gamma_5)q +
\frac{1}{2}(D_{\mu}\pi^a)^+(D_{\mu}\pi^a)+\frac{1}{2}\partial_{\mu}\sigma
\partial^{\mu}\sigma + \frac{1}{2}\mu^2(\sigma^2+\pi^a\pi^a)\nonumber \\
&-\frac{1}{4}\lambda(\sigma^2+\pi^a\pi^a)^2+(D_{\mu}H_A)^+(D_{\mu}H_A)+
\mu_A^2(H_A^+H_A)-\lambda_1(H_A^+H_A)^2
-\lambda_2(H_A^+H_B)(H_B^+H_A)\nonumber \\
&-h(H_A^+H_A)(\sigma^2+\pi^a\pi^a)-\frac{1}{4}B_{\mu \nu}B^{\mu
\nu}-\frac{1}{4}V_{\mu \nu}V^{\mu \nu}-\frac{1}{4}V^a_{\mu
\nu}V_a^{\mu \nu}.
\end{eqnarray}
Here $q = (u,d)$ is the first generation quark doublet (the fundamental representation); $\pi$-triplet is the adjoined representation
of the $SU(2)$ gauge group; $\sigma$-meson is a singlet;
$H_{1,2}$ - two doublets of complex scalar fields with the
hypercharges $Y_{1,2}=\pm 1/2$, $a=1,2,3$ and $A=1,2$. Physical states are formed by the primary fields mixing when
quadratic mass forms of scalar and vector fields are diagonalized.
At the tree level, the mass forms arise as a result of vacuum
shifts which generate the gauge fields masses: $<\sigma>=v,\,\,\, <H_1>=(v_1,0)/\sqrt{2},\,\,\,
<H_2>=(0,v_2)/\sqrt{2}$.
%\begin{equation}
%<\sigma>=v,\,\,\, <H_1>=(v_1,0)/\sqrt{2},\,\,\,
%<H_2>=(0,v_2)/\sqrt{2}. \label{10a}
%\end{equation}

The model Lagrangian in terms of physical fields can be found in \cite{10}.
%is the following:
%\begin{align}\label{E:12a}
% L_{Phys}&=\bar{u}\gamma^{\mu}u(\frac{2}{3}eA_{\mu}+g_{u\omega}\omega_{\mu}+g_{u\rho}\rho^0_{\mu})+
%          \bar{d}\gamma^{\mu}d(-\frac{1}{3}eA_{\mu}+g_{d\omega}\omega_{\mu}+g_{d\rho}\rho^0_{\mu})\notag\\
%          &+ig_2(\pi^{-}\pi^{+}_{,\mu}-\pi^{+}\pi^{-}_{,\mu})(\sin \theta \,\, A^{\mu}-\cos \theta \sin \phi \,\,
%         \omega^{\mu}+\cos \theta \cos\phi \,\,\rho^{0\mu})\notag\\
%         &-\sqrt{2}i\varkappa\pi^{+}\bar{u}\gamma_5 d-\sqrt{2}i\varkappa\pi^{-}\bar{d}\gamma_5 u-i\varkappa\pi^0
%         (\bar{u}\gamma_5 u-\bar{d}\gamma_5 d)\notag\\
%         &+2g_2e\cos \theta \cos\phi \,\,\rho^0_{\mu}A^{\mu}\pi^{+}\pi^{-}-
%          2g_2e\cos \theta \sin\phi
%          \,\,\omega_{\mu}A^{\mu}\pi^{+}\pi^{-}\notag \\
%          &+\frac{1}{\sqrt{2}}g_2 \rho^{+}_{\mu}\bar{u}\gamma^{\mu}d+\frac{1}{\sqrt{2}}g_2 \rho^{-}_{\mu}\bar{d}
%         \gamma^{\mu}u
%          +ig_2\rho^{+\mu}(\pi^0\pi^{-}_{,\mu}-\pi^{-}\pi^0_{,\mu})\notag \\
%          &+ig_2\rho^{-\mu}(\pi^{+}\pi^0_{,\mu}-\pi^0\pi^{+}_{,\mu}).
%\end{align}
Due to universality of vector boson couplings, their interactions with quarks and mesons are described by three gauge parameters: $g_{0}$, $g_{1}$ and $g_{2}$. The Lagrangian vertices contains some orthogonal linear combinations of $g_0, \, g_1, \, g_2$ with (tree) mixing angles $\sin\phi, \,\cos\phi, \,\sin\theta, \, \cos\theta$. These angles are fixed by the experimental data on the vector meson masses and decay widths. Angle $\theta$ is determined by the diagonalization of the vector fields quadratic form. 
%From the mixing it follows that $\omega$- meson is not a pure isoscalar state, however, the isotriplet admixture to $\omega$ is small: it is $ \sim %\sin\phi$ and it is close to zero. So, the corresponding contribution can be omitted in calculations involving the $\omega$- meson in a good %approximation.
%Here $g_{u\omega}, \,g_{u\rho},\, g_{d\omega},\,g_{d\rho}$ - orthogonal linear combinations of $g_0, \, g_1, \, g_2$ %with $\sin\phi, \,\cos\phi, \,\sin\theta, \, \cos\theta$ as (tree) mixing parameters \cite{OURHEP}. Some parameters of %the mixing can be fixed from the experimental data on the vector meson masses and decay widths. Angle $\theta$ is %determined by the
%diagonalization of the vector fields quadratic form. 
%In Eqs.(\ref{E:12a}):
%\begin{align}\label{E:12b}
% &g_{u\omega}=\frac{1}{2}g_1 \cos\phi +\frac{1}{2}\sin\phi \,(\frac{1}{3}g_0\sin \theta-g_2\cos \theta)  \,,\notag\\
% &g_{u\rho}=\frac{1}{2}g_1 \sin\phi -\frac{1}{2}\cos\phi \,(\frac{1}{3}g_0\sin \theta-g_2\cos \theta)\,, \notag\\
% &g_{d\omega}=\frac{1}{2}g_1 \cos\phi +\frac{1}{2} \sin\phi \,(\frac{1}{3}g_0\sin \theta+g_2\cos \theta) \,,\notag\\
% &g_{d\rho}=\frac{1}{2}g_1 \sin\phi -\frac{1}{2} \cos\phi \,(\frac{1}{3}g_0\sin \theta+g_2\cos
% \theta)\,.
%\end{align}

We have the following relations:
 \begin{eqnarray}\label{E:12c}
  \sin \theta=\frac{g_{0}}{\sqrt{g_0^2+g_2^2}},\,\,\,e=g_0\cos \theta,\,\,\, v^2_1+v^2_2=4 \frac{m^2_{\rho^{\pm}}}
  {g^2_2},\,\,\,
  \sin\phi=\frac{g_1}{g_2}\Bigl(\frac{m^2_{\rho^{\pm}}-m^2_{\omega}(g^2_2/g^2_1)}
  {m^2_{\omega}-m^2_{\rho^0}}\Bigr)^{1/2}.
 \end{eqnarray}

 In the model, effective "tree" couplings on the mass shell of external particles are reproduced by the sum of tree and one-loop quark and meson diagrams, they contain some NP scale and depend on the mixing angles, $\phi$ and $\theta$. So, the couplings incorporate both short-distance and long-distance contributions; the "tree" couplings values are fixed from the data on meson two-particle decays with a sufficient accuracy. We suppose, these effective on-shell vertices (not formfactors!) have tensor structure of the bare vertices. From preliminary considerations it follows that extra terms in off-shell formfactors ($\sim k^{\mu}$, where $k$ is the external momenta, for example) contribute into physical quantities (decay widths) no more than $\sim 10 \%$. Thus, firstly, there are loop diagrams, which contribute into on- and off-shell couplings. Approximately, an account of these corrections can be reproduced by some multiplicative parameter in tree amplitudes. Secondly, there are loop diagrams, which cannot be reduced to contributions into the effective couplings and vertex formfactors.  
The obvious difference between effective on-shell constant couplings and off-shell formfactors (depending on the NP cutoff parameter) prevents from the double-counting in the model. Note, in these formfactors the internal particles masses interpret as the physical ones.

Therewith, decays like $\omega \to \pi \gamma$, $\rho \to \pi \gamma$, $\omega, \,\rho \to 3 \pi$, $\sigma \to \gamma \gamma$ and $\pi^0 \to \gamma \gamma$ are described by finite quark loop diagrams. With a good accuracy, the vertex couplings in these diagrams can be taken as the effective $\rho q \bar q$ ($\rho \pi \pi$) tree constants in agreement with the experimental data. Note, these arguments can be applied to the electromagnetic vertices too - the vertices are renormalized by strong interactions, but the corrections from quark and meson loops are small in the case under consideration. This fact is endorsed by calculations of loop processes with on-shell photons, in particular, $\pi^0 \to \gamma \gamma$  (see below). Thus, the consistency and supportability of the approach is verified by the results for all types of possible reactions - both at the tree and the loop level.

To fix the set of the gauge physical couplings and vacuum shifts, we use experimentally observable
mass spectrum and widths of some two-particle hadronic decays of
vector mesons \cite{11} - in our model these processes occur at the tree level, fixing on-shell "tree" couplings. 

 Constants $g_2$, $\sin \phi$ (or $g_1$) and $cos \theta$ are
fixed from the experimental values of $\Gamma
 (\rho^+ \to \pi^+ \pi^0),$ $\Gamma
 (\rho^0 \to \pi^+ \pi^-)$ and  $\Gamma
 (\omega \to \pi^+ \pi^-)$. 
Then, the value of $g_{0}$ can be extracted from the relation $e=g_{0}\cdot
 g_2/(g_{0}^2+g_2^2)^{1/2}$ (see (\ref{E:12c}).

 Thus, from the experimental data and (\ref{E:12c}) we get the set of main parameters values:
 \begin{eqnarray}\label{E:13}
 g_{0}^2/4\pi = 7.32\cdot 10^{-3}, \,\,\,
 g_1^2/4\pi= 2.86, \,\,\, g_2^2/4\pi = 2.81,\nonumber \\
 \sin \phi = 0.031, \, \sin \theta = 0.051,\, v_1^2 + v_2^2\approx
 (250.7 \,\mbox{MeV})^2.
 \end{eqnarray}
 
\section{Radiative decays of vector mesons}

 Due to two-level structure of the model, the meson-meson Lagrangian classifies both tree and loop level processes,
while the quark-meson interactions occur as loop contributions only.

 Radiative decays of vector mesons $\rho^0 \to \pi^+\pi^-\gamma$ and $\omega \to \pi^+\pi^-\gamma$ are some test reactions for the meson-meson sector of the gauge model. Obvious tree diagrams (see \cite{10}) describe these processes with the pair $\pi^+ \pi^-$ production.

Analytical form of the differential width for the process is presented in \cite{10}.
In Fig.1 curve (1) describes the model fit of normalized spectrum of photons,
$dB(E_{\gamma})/dE_{\gamma}=1/ \Gamma_{tot}^{\rho} \cdot
d\Gamma(E_{\gamma})/dE_{\gamma}$, with a single free multiplicative parameter. It improves the description at the photon low
energy range due to an account of $q^2-$ dependence of couplings. Namely, normalizing the
spectrum by the theoretical value $\Gamma_{tot}^{\rho}$, the ratio depends on the $g_{\rho \pi \pi}-$
formfactors at the different energy scales  because intermediate pions are off-shell. Totally, this multiplicative parameter increases the ratio no more than $(8-10)\%$, due to the formfactors contribution. Curve (2) demonstrates comparison of the spectrum  with the experimental data ~\cite{12,13}. To describe the spectrum fine structure
near $E_{max}$, it was suggested to consider the loop corrections (see \cite{14,15}) which do not contribute in vertices formfactors.
%-------------------------------------------------------------
\begin{figure}[h!]
\centerline{\includegraphics[width =.5\textwidth]{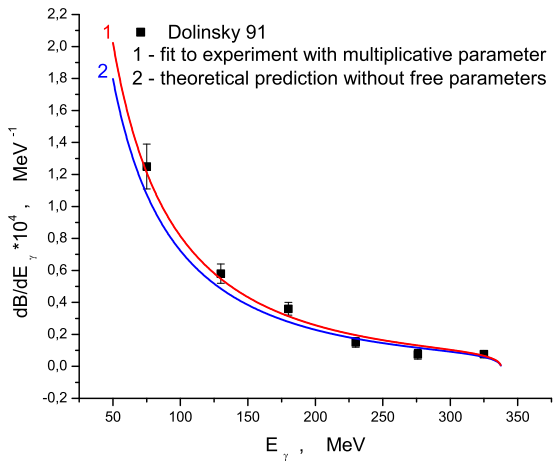}}
\caption{Photons spectrum in $\rho\to 2\pi\gamma$ decay .}
\label{fig:Curve}
\end{figure}
%-------------------------------------------------------------
Integrating of the spectrum from $E_{\gamma}^{min} = 50
\,\mbox{MeV} $ up to $E_{\gamma}^{max} = m_{\rho}(1-4 m_{\pi}^2/m_{\rho}^2)/2$, we get
the value of $\rho$- meson branching $B(\rho^0\to \pi^+
\pi^- \gamma) = 1.17 \cdot 10^{-2}$ which slightly exceeds the
experimental value $B^{exp}(\rho^0\to \pi^+ \pi^- \gamma) = (0.99
\pm 0.16)\cdot 10^{-2}$ \cite{11}. However, an account of loop
contributions (with the phenomenological couplings) leads to even more
result: $B^{phen}(\rho^0\to \pi^+ \pi^- \gamma) = (1.22 \pm
0.02)\cdot 10^{-2}$ ~\cite{14}. An excess of $B^{theor}$ over $B^{exp}$ can be caused by the lack of experimental data at energy range $E_{\gamma}<75
\,\mbox{MeV}$.

Replacing $ \cos\phi \to \sin\phi$ ($\sin\phi \approx 0.034$) and $m_{\rho} \to m_{\omega}$ in the formula for the differential width, we  have $B(\omega\to
\pi^+ \pi^- \gamma) = 4.0 \cdot 10^{-4}$ and $B(\omega\to \pi^+
\pi^- \gamma) = 2.6 \cdot 10^{-4}$ for $E^{min}_{\gamma} = 30 \,
\mbox{MeV}$ and $50\, \mbox{MeV}$, respectively (it agrees with
$B^{exp}(\omega\to \pi^+ \pi^- \gamma) \le 3.6 \cdot 10^{-3}$
~\cite{11}).

Certainly, loop corrections are important and in analogy with ~\cite{14} they can increase
$B(\omega\to \pi^+ \pi^- \gamma)$ up to $(2-3) \cdot 10^{-3}$. Essentially, in the loop level processes all divergencies are summed to zero when all external lines are on the mass shell \cite{16}.

In the quark-meson models, radiative decays $\rho^0,\,
\omega \to \pi^0\gamma$ and three-particle decays $\omega, \rho
\to 3 \pi$ occur via quark loops only with the gauge (tree) vertices. 
Corresponding one-loop diagrams and analytical expressions for the decays $\omega,\, \rho^0 \to
\pi^0\gamma$ are in \cite{10}.  The Goldberger-Treiman relation $\varkappa \approx m_q/f_{\pi}$ defining the constant of
$\sigma q\bar q$ and $\pi q \bar q$ interaction had been used. From this relation it also follows that vacuum shift $v\approx f_{\pi}$. Constituent quark mass value, as a free parameter, is extracted from the widths fit. 
%-------------------------------------------------------------
%\begin{figure}[h!]
%\centerline{\epsfig{file=figure3.eps,width=9cm}}
%\caption{Feynman diagrams for the radiative decay $\rho \to\pi^0
%\gamma$.} \label{fig:Feynm2}
%\end{figure}
%-------------------------------------------------------------

%For  the
%decay $\omega \to \pi^0\gamma$ we have
%\begin{equation}\label{E:5}
%\Gamma(\omega \to \pi^0 \gamma) =\frac{3\alpha
%g_1^2}{2^7\pi^4}\cos^2\phi\,\,
%m_q\frac{m_q^3}{m_{\omega}f_{\pi}^2}\left(1-\frac{m_{\pi}^2}{m_{\omega}^2}\right)
%|L_{\omega}|^2.
%\end{equation}
%Here
%$$L_{\omega}=Li_2\left(\frac{2}{1+\sqrt{\lambda_1}}\right)+Li_2\left(\frac{2}
%{1-\sqrt{\lambda_1}}\right)-Li_2\left(\frac{2}{1+\sqrt{\lambda_2}}\right)
%-Li_2\left(\frac{2}{1-\sqrt{\lambda_2}}\right),$$ and $\lambda_1
%= 1 -4m_q^2/m_{\omega}^2, \, \lambda_2 = 1 -4m_q^2/m_{\pi}^2,$  $f_{\pi} = 93\,\mbox{MeV}.$ $g = g_1e
%\varkappa \cos\phi$ and the Goldberger-Treiman relation $\varkappa \approx m_q/f_{\pi}$ defines the constant of
%$\sigma q\bar q$ and $\pi q \bar
%q$ interaction. Constituent quark mass value, as a free parameter,
%can be found from the widths fit. From this relation it also follows that vacuum shift $v\approx f_{\pi}$.
 
The widths of $\rho^0,\,
\omega \to \pi^0\gamma$ agree with the experimental data \cite{11} if we take $m_q = 175 \pm 5 \,\mbox{MeV}$ and keep an imaginary part of the amplitude as nonzero:
\begin{eqnarray}\label{E:7}
 \Gamma^{theor}(\omega \to \pi^0\gamma)=0.74\pm
0.02 \, \mbox{MeV},\quad \quad \Gamma^{exp}(\omega
\to \pi^0\gamma)=0.76 \pm 0.02 \,\mbox{MeV};\nonumber \\
\Gamma^{theor}(\rho^0 \to \pi^0\gamma) = 0.081\pm 0.003 \, \mbox{MeV},
\quad \quad \Gamma^{exp}(\rho^0 \to \pi^0\gamma)=0.090 \pm 0.012 \,
\mbox{MeV}.
\end{eqnarray}
Note, isotopic structure of vertices gives: $\Gamma(\rho^0 \to \pi^0\gamma)/\Gamma(\omega \to \pi^0\gamma) \approx 1/3^2$, and it is confirmed by the results of calculations.

Taking the small value $m_q =(170-180)\, \mbox{MeV}$, we assume that the (constituent) quarks propagate and annihilate in the internal NP hadron vacuum, so they are not free asymptotic states. I.e., the imaginary parts of amplitudes describe the constituent quark propagation up to the distances, where the NP effects are sufficiently large to bind the quarks into a real meson. Attractive forces between the quarks lead to the effective decreasing of their mass.
Instead, known "naive" idea can be used: imaginary part of quark loop equals zero "by hand" \cite{17,18}. Then, from the above widths we get $m_q = 280 - 290 \,\mbox{MeV}$.
A further way to provide the "infrared confinement" was suggested in \cite{19,20} by the introducing of a special cutoff parameter, $\lambda$, as a common scale in $\alpha-$ representation of the process amplitude. In particular, from this procedure the decay widths follow in a good agreement with the data, if $m_q = 280 - 290 \,\mbox{MeV}$ and $\lambda = 260 \,\mbox{MeV}$. At the same time, the value $\lambda = 210 \,\mbox{MeV}$ leads to light quarks, $m_q = 170 - 180 \,\mbox{MeV}$ and nearly the same widths values. In all cases, the ratio $\lambda/m_q =0.9 - 1.2$.
 
 \section{Effective vertices, two-gamma decays and scalar mesons}
 
As it is known, decay $\pi^0 \to \gamma \gamma$ is described by finite triangle quark diagrams. The decay width is:
\begin{eqnarray} \label{E:15a}
\Gamma(\pi^0\to \gamma \gamma)=\frac{\alpha^2 \varkappa^2}{4
\pi^3} \frac{m_q^2}{m_{\pi}}(\arcsin \frac{m_{\pi}}{2m_q})^4.
\end{eqnarray}
%Here $C_0(0,0,m_{\pi}^2;m_q,m_q,m_q)-$ is a special case of the scalar three-point
%Passarino-Veltman function, which is reduced to the form
%$$
%C_0(0,0,m_{\pi}^2;m_q,m_q,m_q)=
%\frac{1}{m_{\pi}^2}[Li_2(\frac{2}{1+\sqrt{\lambda}})+Li_2(\frac{2}{1-\sqrt{\lambda}})] = \frac{2}{m_{\pi}^2}[\arcsin \frac{m_{\pi}}{2m_q}]^2,
%$$
%where $\lambda=1-4m_q^2/m_{\pi}^2$. 
Numerically, $\Gamma(\pi^0\to \gamma \gamma)=8.48 \, \mbox{eV}$
for $m_q=175\,\mbox{MeV}$. If $m_q=300 \,
\mbox{MeV}$, the width is $\Gamma(\pi^0\to \gamma \gamma)=7.91 \,
\mbox{eV}$. In the exact chiral limit \cite{21} $\Gamma(\pi^0\to \gamma \gamma)=7.63 \, \mbox{eV}$, and the experimental value is in the interval $7.22 \, \mbox{eV}\le \Gamma(\pi^0\to \gamma
\gamma)\le 8.33 \, \mbox{eV}$ \cite{11}. However, it should be taken into account: a) a possible violation of the Goldberger-Treiman relation can be as much as $(3-4) \%$ \cite{22} (it means
a weak $q^2-$ dependence of $\varkappa$); b) $\pi^0 - \eta-$ mixing can be noticeable for
the case  \cite{23}, c) loop corrections to
$\gamma q \bar q$ vertex are also significant. Namely, taking $m_q = 175 \, \mbox{MeV}$ and decreasing effective $\pi q \bar q$ coupling by $4 \%$ only, we get $\Gamma(\pi^0\to \gamma \gamma)=7.82 \,
\mbox{eV}$. To explain the discrepance with the data, it is sufficient also to
decrease the effective electromagnetic coupling  $\gamma q \bar q$ no more than $2\%$ (it can be provided by  small variation of NP scale, $\Lambda$, in the diverged triangles of $q \sigma q$ and $\pi q \pi$ types). Moreover, for the decay amplitude imaginary part equals zero. Nevertheless, introducing an infrared cutoff with $\lambda = 100\, \mbox{MeV}$, for $m_q = 175 \, \mbox{MeV}$ we have $\Gamma(\pi^0\to \gamma \gamma)=7.63 \,\mbox{eV}$. However, nonuniversality of the parameter $\lambda$ arises in the case. If $m_q\sim 300\,\mbox{MeV}$, from exact integration of the amplitude we get a good agreement with the data.

From quark loops an effective couplings can be determined. Then, $g_{\rho \omega \pi}^{exp} = (15-17) \, GeV^{-1}$ and the model gives 
$g_{\rho \omega \pi}^{theor} = 14.5 \, GeV^{-1}$ if $m_q=175 \,\mbox{MeV},\, \lambda=250 \,\mbox{MeV}$;  $g_{\rho \omega \pi}^{theor} = 17.7 \, GeV^{-1}$ for $m_q=280 \,\mbox{MeV},\, \lambda=280 \,\mbox{MeV}$ and $g_{\rho \omega \pi}^{theor} = 16.4 \, GeV^{-1}$ when $m_q=300 \,\mbox{MeV}, \,\lambda=280\, \mbox{MeV}$.

Analogously, $g_{\rho \pi \gamma}^{exp} = 0.723 \pm 0.037 \, GeV^{-1}$  and $g_{\rho \pi \gamma}^{theor} = 0.743 \, GeV^{-1}$ when $m_q=280\, \mbox{MeV},\, \lambda=280\, \mbox{MeV}$; $g_{\pi \gamma \gamma}^{exp} = 0.276 \, GeV^{-1}$  and $g_{\pi \gamma \gamma}^{theor} = 0.278 \, GeV^{-1}\, m_q=280 \,\mbox{MeV}$.
For all these parameters we get the ratio $\lambda /m_q = 0.9 - 1.2$, excepting light quark case with $m_q=175 \,\mbox{MeV}$.

Scalar isotriplet $a_0(980)$ occurs in the model with an accuracy up to small mixing angles, $\phi$ and
$\theta$. Namely, from initial Higgs doublets with $Y=\pm 1/2$ after the
spontaneous breaking, two charged scalar combinations arise. They can
be unified with the residual neutral components to
constitute the isotriplet. In the simple case of zero mixing (then exact isotriplet structure emerges and $v_1=v_2$, but this equality should be slightly broken by electromagnetic forces), $a\omega \rho$ and $aa\rho$ parts of the Lagrangian have the form, where isovectors $(\rho^+,\rho^0,\rho^-)$ and $(a^+,a^0,a^-)$ are introduced: 
\begin{eqnarray}\label{E:14a}
L_{a\omega \rho} &= \frac{1}{\sqrt{2}}\,v g_2 g_3\,\omega^{\mu}(a^-
\rho^{+}_{\mu}+ a^+ \rho^{-}_{\mu}+a^0 \rho^{0}_{\mu})=
\frac{1}{\sqrt{2}}\,v g_2
g_3\,\omega_{\mu}\rho^{\mu}_{\alpha}a_{\alpha},\nonumber \\
L_{aa\rho} &= \frac{i}{2}\, g_2\, [\rho^{0\mu} (\partial_{\mu}a^+
\cdot a^- -\partial_{\mu}a^- \cdot a^+)+\rho^{+\mu}
(\partial_{\mu}a^-
\cdot a^0 -\partial_{\mu}a^0 \cdot a^-)\nonumber \\
&+\rho^{-\mu} (\partial_{\mu}a^0 \cdot a^+ -\partial_{\mu}a^+ \cdot
a^0)]= \frac{i}{2}\,
g_2\,\epsilon_{\alpha\beta\gamma}\rho^{\mu}_{\alpha}a_{\beta}\partial_{\mu}a_{\gamma}.
\end{eqnarray}

Interaction of scalars with $\pi$-mesons is described by the Lagrangian
 \begin{eqnarray}\label{E:14}
  L_{\pi h}=(\pi^0\pi^0+2\pi^{+}\pi^{-})(g_{\sigma\pi}\sigma_0+g_{f\pi}f_0+g_{a\pi}a_0).
 \end{eqnarray}
 
 There is an interesting feature of this sector: mixing angle $\psi$ defining the physical fields $\sigma = \cos \psi \cdot \sigma_0 + \sin \psi \cdot f_0, \, f_0 = -\sin \psi \cdot \sigma_0 + \cos \psi \cdot f_0$ can be equaled zero. Then, from explicit expressions $\sin \psi \sim (m_{f_0}^2-a-b)=0$, $m_{a_0}=a-b$ we found:
 $m_{f_0}= m_{a_0}\approx hv^2-\mu_1^2+3\lambda_2v_1^2$, if $b=2\lambda_1v_1^2 \approx 0$. It means $\lambda_1 \approx 0$, so the term $\lambda_1(H_A^+H_A)^2$ in the initial Lagrangian is damped. From experimentally known (small) difference of $a_0$ and $f_0$ masses an estimation of $\lambda_1$ can be extracted, because $\Delta m = 4\lambda_1v_1^2$. Also, from $\sin \psi = 0$ it follows $g_{\sigma \rho \rho} =0$; as a consequence, scalars $f_0, \, a_0$ do not interact with quark current directly.  
 
 Due to free parameters, it is possible to consider dominant decay
 channels of scalar mesons -  $f_0(980)\to\pi\pi$ and $\sigma_0\equiv f_0(600)\to \pi\pi$ ~\cite{11}.
Decay $a_0\to\pi\pi$ is not observed, so from $g_{a\pi}=h(v_2-v_1)/\sqrt{2}$ it follows that $v_2\approxeq v_1$.
Exact symmetry, $v_1=v_2$, is not possible because it forbids the suppressed (but observed) decay $\omega\to\pi^{+}\pi^{-}$.
  
At the tree level two-photon decays of scalar mesons are absent. Considering $\sigma \to \gamma \gamma$ decay, we have triangle loop diagrams mediated by quarks, pions, $a_0$- and $\rho$-mesons. Due to above arguments, vector meson loops are absent. Total amplitude of the process is gauge invariant without any divergencies: $M_{tot}(\sigma \to \gamma \gamma)\sim (g_{\mu \nu}\cdot k_1k_2 -k_{1,\mu}k_{2,\nu})\cdot [M_{q-loop}+M_{\pi-loop}+M_{a_0-loop}].$ Experimentally, $\Gamma^{exp}(\sigma \to \gamma \gamma)=(1-5)\, \mbox{KeV}$ \cite{11}. From the "naive approximation" we get $\Gamma^{theor}(\sigma \to \gamma \gamma)=(3-5)\, \mbox{KeV}$ for $m_q=280 \, \mbox{MeV}, \,\, m_{\sigma}=450 \pm 50 \,\mbox{Mev}$ and $g_{\sigma \pi \pi}= (m_{\sigma}^2-m_{\pi}^2)/2f_{\pi}$ beyond the chiral limit. Besides, $\Gamma^{theor}(\sigma \to \gamma \gamma)=(2.5-5)\, \mbox{KeV}$ for $m_q=175 \, \mbox{MeV}$; in this case an agreement can be achieved only if $g_{\sigma \pi \pi}=(0.5 - 1.5)$ of the chiral limit value and $m_{\sigma}=700 \pm 100 \,\mbox{Mev}$. This value contradicts to known data $m_{\sigma}= 400-500 \,\mbox{MeV}$. Note, imaginary part of $M_{tot}$ is small for all cases, and $g_{\sigma a_0 a_0} \approx 1/3 g_{\sigma \pi \pi}$. 

Now, the "infrared confinement" procedure results to $\Gamma^{theor}(\sigma \to \gamma \gamma)=(1.5-2.5)\, \mbox{KeV}$ for $m_q=280 \, \mbox{MeV},\, m_{\sigma}=500\pm 50 \,\mbox{MeV}$ and $\lambda =260\, \mbox{MeV}$. All these values provide good agreement with the data for $\sigma \to\pi\pi$ and $f_0(980)\to\pi\pi$ decays. As to $a_0 \to \gamma \gamma$ and $f_0 \to \gamma \gamma$, it has been found in \cite{24} that the decays can be explained only by four-(constituent) quark components of the mesons. In the gauge model, these components are simulated by vector meson (due to $a_0 \rho \omega, \, f_0 \rho \rho$ vertices) and scalar meson loops (due to $f_0 a_0 a_0, \, f_0 \pi \pi$ verices); analysis of these decays is in progress. 

To include axial vector mesons and strange quarks, the model should be generalized to chiral $SU(3) \times SU(3)$. Then dominant decay channel of the scalar mesons ($a_0 \to K \bar K$ decay, for example) will be understood and calculable as a possible admixture of $K \bar K-$ states into scalar structure. Note, however, the decay $a_0(980) \to \eta \pi$ can be described by the symmetry allowed extra term $\sim \eta \pi_a H^+_A \tau^a H_A$ in the model Lagrangian.

\section{Conclusions}
The gauge quantum field approach is applied to low-energy hadron interactions using the Higgs mechanism in a framework of $\sigma-$ model with the VMD. Residual Higgs degrees of freedom allow to reproduce $a_0(980)$ and $f_0(980)$ scalars. Due to free parameters, the mass degeneration of these mesons is explained. Decay $a_0 \to \pi \pi$ is damped, as it is known from the expriment, two-pion decays of $f_0-$ ans $\sigma-$ mesons agree with the experimental data well. We also conclude that the two-quark components of scalar mesons ($a_0, \, f_0$) are small - they occur at the loop level only.

Radiative decays of vector mesons (at the tree and loop level) and two-gamma decays of $\sigma-$ and $\pi^0-$ mesons are well described too, the values $m_q=280 \, \mbox{MeV}$ and $m_{\sigma} = (450 - 550)\, \mbox{MeV}$ are preferred by the analysis with the "infrared confinement" procedure. Also, the values of effective vertices $g_{\rho \omega \pi}, \, g_{\rho \pi \gamma}, \, g_{\pi \gamma \gamma}$ correspond to known experimental data. 

So, the gauge renormalizable generalization of $\sigma-$ model based on $U_0(1) \times U(1) \times SU(2)$ group with the spontaneous symmetry breaking and the VMD is a reasonable and consequtive way to consider radiative and hadronic decays of vector and scalar mesons. Some important features of scalar mesons spectrum are explained in the model due to interpretation of these mesons as vacuum fields. Chiral generalization of the model should be done to include axial vector states and other scalar mesons into the gauge scheme.

\end{document}